\documentclass[graybox]{svmult}

\usepackage{type1cm}        
\usepackage{makeidx}         
\usepackage{graphicx}        
\usepackage{multicol}        
\usepackage[bottom]{footmisc}

\usepackage{newtxtext}       %
\usepackage[varvw]{newtxmath}       

\newcommand{\beq}{\begin{equation}}
\newcommand{\eeq}{\end{equation}}
\newcommand{\barr}{\begin{eqnarray}}
\newcommand{\earr}{\end{eqnarray}}
\newcommand{\TM}{T_{\textrm{M}}}
\newcommand{\TCMB}{T_{\textrm{CMB}}}
\newcommand{\dd}{\mathrm{d}}

\makeindex             

\begin{document}

\title*{CMB and accretion}
\author{Pasquale Dario Serpico}
\institute{LAPTh, CNRS and Univ. of Savoie Mont Blanc, 9 chemin de Bellevue - BP 110, 74941 Annecy, France, \email{serpico@lapth.cnrs.fr}
}
%
%
\maketitle

\abstract{Primordial black holes (PBHs) of stellar mass or heavier would  accrete baryonic gas, which becomes denser and hotter, injecting energetic photons in the cosmological medium soon after cosmic recombination, in the so-called {\it dark ages}. The ionisation history of the universe would be altered, an effect cosmic microwave background (CMB) temperature and polarisation anisotropies is sensitive to. The magnitude of the effect depends on the abundance and mass distribution of the PBHs, as well as on still uncertain aspects of accretion luminosity. We review the current understanding of this field, with an emphasis on the peculiarities of the phenomenon in the cosmological context, and present the existing constraints on the PBH abundances.}

\section{Introduction}\label{sec:1}
Massive bodies gravitationally accrete matter, which acquires kinetic energy at the expense of potential one; as a result of microscopic dissipative mechanisms, the accreted material  heats up and emits radiation. At the same time,  following  energy and angular momentum losses, this material falls onto the massive body or, for a black hole (BH), it crosses the event horizon. A rich astrophysical phenomenology is associated to accretion power (see e.g. the monograph~\cite{2002apa..book.....F}). The luminosity due to the accretion of the interstellar medium by isolated stellar or planetary objects, while universal, is typically too modest to be of observational relevance. This is why accretion studies in astrophysics often focus on close binaries or episodes of large mass accretion, for instance related to a stellar disruption by a supermassive black hole (SMBH). Most electromagnetic signals associated to BH detection are related to these situations. 

Needless to say,  accretion phenomena on compact objects require the presence of compact objects in the first place. In the standard cosmological scenario, these are only present starting from the end of the dark ages, let us say at redshifts $z\lesssim 20-30$, after first stellar structures could form by gravitational collapse in the pristine dark matter (DM) halos. While differing in poorly known quantitative details, the  accretion phenomena in these environments are qualitatively similar to the more familiar ones going on in astrophysical settings at low $z$.

 In scenarios where sufficiently massive primordial black holes (PBHs) are present, however, it becomes of interest to study the relevance of accretion phenomena onto them in the early universe, at $z\gg 20$. In this case, the direct detection of the accretion luminosity may be impossible or impractical. Currently, the most effective tool to constrain such an exotic process is by studying the alterations it provokes on the cosmic microwave background (CMB) temperature and polarisation anisotropies. This chapter is devoted to outline the key ingredients, equations and assumptions entering these calculations, and to present current constraints to an audience presumably already familiar with the basic of the standard $\Lambda$CDM cosmology. The chapter is structured as follows: In Sec.~\ref{EinjCMB} we recall the key arguments on why and how CMB anisotropies can be affected by such energy injection.  Sec.~\ref{AccRateLum} reviews the basic notions on accretion rate and luminosity, while Sec.~\ref{AccCosmo} specialises these notions to a cosmological setting. Sec.~\ref{Disks} discusses {\it disk} accretion and luminosity in general and in the cosmological context, and Sec.~\ref{DMhalo} shows how to account for the presence of a DM halo around the accreting PBH. The latter treatment does not depend on the nature of DM, under the sole hypothesis that the bulk of DM is collisionless. Sec.~\ref{ConstrDisc} presents current constraints and a discussion on some perspectives.  Natural units are used in equations, while numerical values are quoted in the most widely used astrophysical units.

\section{Impact of energy injection on CMB anisotropies}\label{EinjCMB}
In the last couple of decades, CMB  anisotropies have been repeatedly used to constrain the energy injection associated to a number of exotic processes; we just mention here seminal articles related to decaying relics~\cite{Chen:2003gz}, annihilating relics like weakly interacting massive particle dark matter~\cite{Slatyer:2009yq,Cirelli:2009bb}, evaporating (hence ``light'') PBHs~\cite{Carr:2009jm} or accreting (hence ``heavy'') PBHs~\cite{Ricotti:2007au}. These putative processes have all qualitatively similar effects on the CMB, reviewed in this section; for details see e.g.~\cite{Poulin:2016anj}.

Note that the energy of the injected particles, even if negligible with respect to the large CMB photon energy density, is not negligible with respect to the kinetic energy of the baryonic gas, which is cold and mostly neutral during the dark ages ($20\lesssim z\lesssim 1000$). 
 The main consequence of this injection is to modify the fraction of free electrons $x_e$, either  directly, via ionisation, or in a multi-step process, via collisional excitation followed by photoionisation by a CMB photon. An indirect effect is the heating of the medium, whose temperature $\TM$ has a feedback on the evolution of $x_e$. CMB anisotropies are sensitive to these effects via alterations to optical depth $\tau$ (integrated value of $x_e(z)$ over redshift) and its time dependence (or ``visibility function'').
 
Modern treatments of the relevant physics are quite detailed and involve numerous processes and transitions, see for instance refs.~\cite{Seager:1999bc,Ali-Haimoud:2010hou} at the basis of widely used codes.  The key  physics can be however understood within the so-called ``Peebles recombination'' model~\cite{Peebles:1968ja,Zeldovich:1968qba}, where these quantities are ruled by a system of two coupled differential equations  (for details on these terms and a compact overview, we refer the reader to the appendix A of Ref.~\cite{Poulin:2015pna}): 
\begin{eqnarray}\label{eq:x_e&T_M}
\frac{\dd x_{e}(z)}{\dd z}&=&\frac{1}{(1+z)H(z)}(R(z)-I(z)-I_X(z))~,\nonumber\\
\frac{\dd\TM}{\dd z}  &=&  \frac{1}{1+z}\bigg[2\TM+\beta_{\rm cool}(\TM-\TCMB)\bigg]+K_h~.
\end{eqnarray}
Here, $H(z)$ is the Hubble expansion rate (dominated by matter in the relevant redshift interval) the $R$ and $I$ terms are the standard recombination and ionisation rates, respectively, while $\beta_{\rm cool}\equiv 8\sigma_T u_{\rm CMB} x_e/[3(1+z)(1+x_e)H\,m_e]$ is the so-called opacity of the gas, with $u_{\rm CMB}=\pi^2 T_{\rm CMB}^4/15$ the energy density of the CMB, $\sigma_T$ the Thomson cross section, $m_e$ the electron mass.   The exotic terms are
represented by the effective ionisation rate  $I_X(z) = I_{Xi}(z)+I_{X\alpha}(z)$ (contributed to by the direct ionisation rate $I_{Xi}$ and excitation+ionisation rate $I_{X{\alpha}}$) and $K_h$, which describes the heating rate due to the new process. All these quantities, standard and extra terms, are in general known (non-linear) functions of $x_e,$ and $\TM$. The extra terms are proportional
 to the deposited energy rate in each channel,  
 \begin{equation}\label{eq:Kh}
\frac{\dd E}{\dd V\dd t}\bigg |_{\textrm{dep}, h}\quad,
\end{equation}
which are related to the direct input of the extra injected energy rates, $ \dd E/(\dd V\dd t)\big|_{\rm inj}$, via transfer functions $f_c$ specific to the channel $c$,
 \begin{equation}
f_c(z,x_{\rm e})  \equiv  \frac{\dd E/(\dd V \dd t)\big|_{{\rm dep},c}}{\dd E/(\dd V\dd t)\big|_{\rm inj}}\,.\label{fzexpr}
\end{equation}
We refer to appendix A of Ref.~\cite{Poulin:2015pna} for further definitions and more details on each coefficient.

The  injection energy rates are the direct input of the model being tested. For instance, for PBHs constituting a fraction  $f_{\rm PBH}$ of the DM, one has
 \begin{equation}
\left.\frac{\dd E}{\dd V \dd t}\right|_{\rm inj}=L n_{\rm pbh}=L f_{\rm PBH}\frac{\rho_{\rm DM}}{M}
\end{equation}
where we are implicitly assuming a monochromatic mass spectrum at $M$ (see Sec.~\ref{ConstrDisc} for a generalisation), and $L$ is the associated accretion luminosity, detailed in the next sections. On the contrary, the transfer functions depend only on standard physics, but their calculation may be rather cumbersome, requiring to follow the degradation of energy and cascades of particle multiplication which are generally non-local (a particle injected at $z_1$ may deposit some of its energy at $z_2<z_1$). Several calculations have been performed in the literature, see e.g.~\cite{Slatyer:2009yq,Valdes:2009cq}  for early ones.
Nowadays, dedicated software exists like the \texttt{EXOCLASS}  package~\cite{Stocker:2018avm}, based on the more modern calculations of~\cite{Slatyer:2015kla},  which can be directly interfaced with the Boltzmann code \texttt{CLASS}~\cite{Blas:2011rf} to evaluate the impact of any kind of exotic electromagnetic energy injection onto CMB anisotropies.
While the efforts towards more realistic treatment of energy deposition is still ongoing, see~\cite{Liu:2023fgu,Liu:2023nct}, it is fair to say that remaining uncertainties are not the dominant ones for the problem of interest here (redshift range $100 \lesssim z\lesssim 1000$, quasi-homogeneous medium). Also note that, while conceptually the problem may be complicated by a feedback effect of the deposited energy on the  deposition function itself, within current constraints this effect is at the percent level, see e.g.~\cite{Liu:2023nct}.

\begin{figure}
\centering
\includegraphics[scale=0.35]{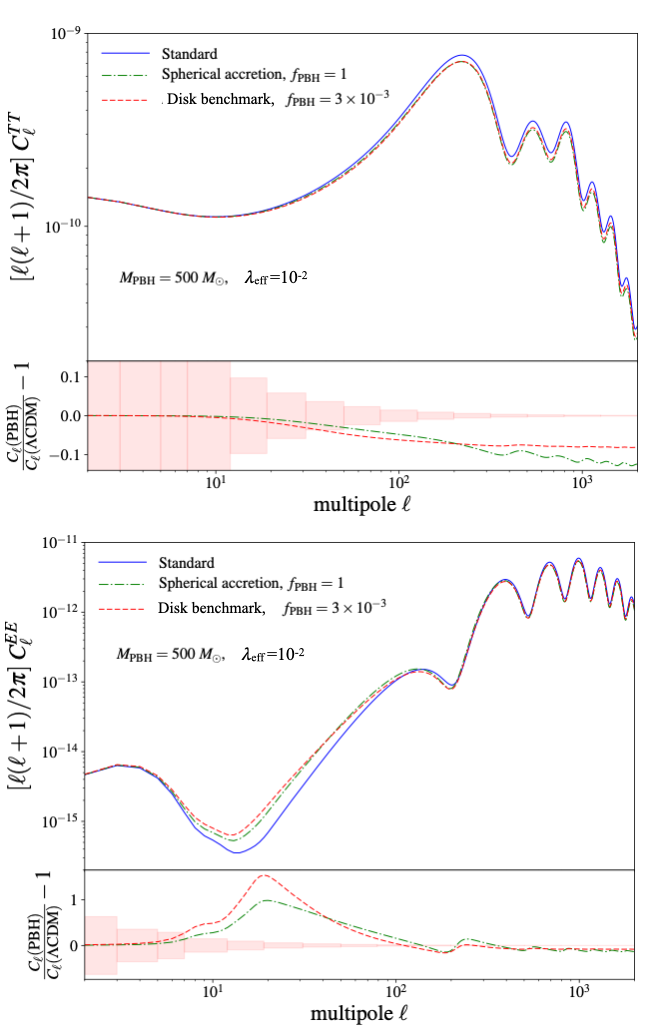}
\caption{CMB temperature (TT, top panel) and polarization  (EE mode, bottom panel) angular power spectrum obtained for a monochromatic population of PBH with masses 500 $M_\odot$ depending on the accretion recipe used. See Sect.s~\ref{AccCosmo},~\ref{Disks} for a description of the two reported benchmarks.
Adapted from Ref.~\cite{Poulin:2017bwe}.} \label{fig:Cl}
\end{figure}

Once the function $x_e(z)$ is known, the impact on the CMB anisotropy angular power spectra is due to the two  previously mentioned effects:
 \begin{itemize} 
  \item The visibility function is altered and recombination is delayed, which  (slightly) shifts the acoustic peaks and causes small wiggles at high multipoles $\ell$ in the differences with respect to a standard $\Lambda$CDM scenario. 
 \item Due to overall higher integrated ionisation fraction, the optical depth is increased (all other parameters being the same). This manifests itself in a damping of temperature anisotropies and an enhanced power in the polarisation spectrum at low/intermediate $\ell$'s. 
 \end{itemize}
 These effects, quantitatively depending on the accretion luminosity evolution, are illustrated in Fig.~\ref{fig:Cl}, taken from~\cite{Poulin:2017bwe}, for two benchmark cases motivated in Sect.s~\ref{AccCosmo},~\ref{Disks}. In particular, the former effect is  visible in the top panel, the latter effect in the bottom panel.
 
Note that in principle, different exotic injection histories due to different processes or different accretion luminosity evolutions correspond to different $x_e(z)$ (see green vs. red lines in Fig.~\ref{fig:Cl}) and could be distinguished via a fine analysis of CMB anisotropies. 
Hence, if a signal were found,  some constraints could be put on the PBH accretion mechanism. With current precision, however, only a rough sensitivity to extreme differences in $x_e(z)$ can be expected, and we will not indulge in these aspects any further.

\section{Generalities on accretion rate and luminosity}
\label{AccRateLum}

A particle of mass $m$ falling without losses from rest at infinity to a distance $R$  from a point mass $M$ acquires kinetic energy equal to the potential one, $G Mm/R=0.5 m R_S/R$, where we introduced the Schwarzschild radius $R_S\equiv 2GM$. Since some of this energy ends up in radiation, typically via collisional/dissipative effects, an electromagnetic luminosity $L$ is associated to accretion. The total electromagnetic luminosity $L$ associated to mass accretion rate $\dot M$ is customarily  written as $L=\epsilon \dot M$, keeping in mind that the conversion efficiency parameter $\epsilon$ is a function of $\dot M$ and of the BH and medium properties. 
If all the kinetic energy were converted into radiation,  from the previous argument one would estimate $\epsilon\leq R_S/(2R)\leq 0.5$. The  part of the mass accretion rate contributing to increase the BH mass hence writes $(1-\epsilon)\dot M$. 
From a macroscopic point of view, the  problem is to determine  $\epsilon$ and $\dot M$.
 
 As a rule of thumb, only particles orbiting at or beyond the innermost stable circular orbit (ISCO) have time for radiative processes to be relevant: Once at the ISCO, the particle crosses the event horizon within a free-fall time~\cite{2002apa..book.....F}.   The closest the ISCO, the higher is $\epsilon$: A typical benchmark value used in the literature is
 $\epsilon\simeq 0.1$ for a Schwarzschild BH, for which $R_{\rm ISCO}=3\,R_S$ (a slightly lower $\epsilon$ is actually expected when taking GR effects into account). For comparison,  $\epsilon \simeq 0.4$ for a maximally rotating Kerr BH. 
 
It is also worth stressing that there are physical limitations to the electromagnetic luminosity.  The most famous such bound is the Eddington limit, when radiation pressure in a spherical setting  prevents  further steady-state accretion. In terms of the proton mass $m_p$ and the Thomson cross section $\sigma_T$,  for a hydrogen gas this condition writes $L_{\rm Edd}=4\pi G Mm_p /\sigma_T\simeq 1.3\times 10^{38} (M/M_\odot)\,$ erg/s. The Eddington limit suggests that: a) With growing BH mass, a proportionally larger limiting luminosity can be attained. b)
Steady-state accretion at larger and larger $\dot M$ is only possible at the expense of a decreasing efficiency $\epsilon$, otherwise the bound would be violated. 
 
Concerning $\dot M$,  in most astrophysical settings of interest it is set  either by the close binary system specification, or is an external input entering e.g. a fit to the data. For PBHs, we are however interested in computing it in a quasi-homogeneous cosmological background.   Even the deceivingly simple problem of homogeneous accretion onto a point mass $M$, despite being rather old, does not have a general solution. Under steady-state hypothesis, the problem was solved by Hoyle \& Littleton in 1939-40~\cite{1939PCPS...35..405H,1939PCPS...35..592H,1940PCPS...36..325H} for an  infinite and pressureless gas cloud with density $\rho_\infty $ far away from the BH, moving at $v_{\rm rel}$,  yielding
 \begin{equation}
 \dot M_{HL}=4\pi \rho_\infty \frac{(G M)^2}{v_{\rm rel}^3}\label{HLAcc}\,.
 \end{equation}
  As shown by Bondi and Hoyle in 1944~\cite{Bondi:1944jm}, up to a factor two smaller accretion is estimated once accounting for density inhomogeneities and the accretion wake.
In the opposite limit of accretion at rest in gas with pressure,  Bondi
found in 1952~\cite{Bondi:1952ni} (see also~\cite{2002apa..book.....F} for a modern exposition) that
   \begin{equation}
 \dot M_{B}=4\pi \lambda \rho_\infty \frac{(G M)^2}{c_{s,\infty}^3}\,,\label{BondiAcc}
 \end{equation}
 where $c_{s,\infty}$ is the speed of sound in the homogeneous gas at infinity and $\lambda\sim {\cal O}(0.1-1)$ is the accretion eigenvalue: It is the key  output of the solution of the steady-state problem involving mass conservation, Euler equation, energy conservation and the equation of state (EOS). We will see a generalisation of this problem in the cosmological framework in the next section. Note that the Bondi problem (or its relativistic generalisation~\cite{1972Ap&SS..15..153M}) yields a transonic solution for BHs, i.e. the velocity changes from subsonic at large distances to supersonic below a finite distance above the event horizon.

We can summarise both regimes of eq.s~\eqref{HLAcc},\eqref{BondiAcc} and provide an interpolation between the two via the following parameterisation:
 \begin{equation}
 \dot M=4\pi \lambda_{\rm eff} \rho_\infty v_{\rm rel}r_{\rm eff}^2\,,  \label{genAcc}
 \end{equation}
 where 
     \begin{equation}
 r_{\rm eff}=\frac{G M}{v_{\rm eff}^2 }\,,\label{reff}
 \end{equation}
 where  $v_{\rm eff}\simeq\sqrt{c_{s,\infty}^2+v_{\rm PBH}^2}$, as suggested by Bondi himself~\cite{Bondi:1952ni} , and $\lambda_{\rm eff}\sim 0.1-1$ is to be adjusted to  the problem at hand.  The variable $ r_{\rm eff}$ thus defined is inspired by the so-called Bondi radius $r_B\equiv GM/c_{s,\infty}^2$; we see that the limit $v_{\rm eff}\to c_{s,\infty}$ in eq.~\eqref{genAcc} yields indeed eq.~\eqref{BondiAcc}. Note that none of the solutions corresponding to eq.s~\eqref{HLAcc},\eqref{BondiAcc} accounts for the radiation feedback onto the flow. 
Also,  eq.s~\eqref{HLAcc},\eqref{BondiAcc},\eqref{genAcc}  grow quadratically with the mass. In general, one does find that  $\dot M/M$ is a growing function of the mass: As a corollary, accretion power only matters for sufficiently heavy objects, typically of stellar size or heavier. 

 The density and temperature profiles corresponding to a given solution can be used  to compute the radiation emitted by the system, accounting for the relevant processes, such a Bremsstrahlung radiation, synchrotron radiation, etc. Once the energy-integrated emissivity (energy per unit time and volume) $j$  is known, the luminosity can be obtained by integration, e.g. in case of spherical symmetry one has $L=\int \dd r 4\pi r^2 j$. Energy-differential quantities can also be derived, of course, if needed.

\section{Accretion in a cosmological setting}\label{AccCosmo}

The first seminal studies of the PBH accretion in a cosmological context, with application to the CMB, are the articles~\cite{Ricotti:2007au,Ricotti:2007jk}. While some of the results of these works remain correct, other results have been proven incorrect and amended in~\cite{Ali-Haimoud:2016mbv}, on which the following discussion is mostly based.

First, the Bondi problem previously introduced can be generalised to include energy cooling and momentum drag via Compton scattering (quantified via the functions $\beta_{\rm cool}$ 
and $\beta_{\rm drag}\equiv (1+x_e) m_e\beta_{\rm cool}/(2m_p) \ll \beta_{\rm cool}$, respectively, named  $\gamma$ and $\beta$ in~\cite{Ali-Haimoud:2016mbv}). 
In spherical symmetry, the relevant system of equations, i.e. mass conservation, momentum equation,
 heat equation and perfect gas EOS write
\barr
4 \pi r^2 \rho |v| &=& \dot{M} = \textrm{const}=4 \lambda \pi  \rho_\infty \frac{(GM)^2}{c_{s,\infty}^3} ,\\
v \frac{\dd v}{\dd r} &=& - \frac{G M}{r^2}  - \frac1{\rho} \frac{\dd P}{\dd r} - \beta_{\rm drag} v, \label{eq:momentum}\\
v \rho^{2/3} \frac{\dd}{\dd r}\left( \frac{T}{\rho^{2/3}}\right) &= &\beta_{\rm cool}(T_{\rm CMB} - T), \label{eq:heat-compt}\\
P &=& \frac{\rho}{m_p}(1 + \overline{x}_e) T,
\earr
for a PBH accreting hydrogen gas at rest in the cosmic frame, where the first mass-conservation equation introduces the accretion eigenvalue $\lambda$ and links the accretion to asymptotic (cosmological homogeneous) density $\rho_\infty$ and sound speed  $c_{s,\infty}=\sqrt{\gamma P_\infty/\rho_\infty}$. The value of the EOS parameter are $\gamma=5/3$ for adiabatic conditions and $\gamma=1$ for isothermal ones.
Remarkably, the authors of~\cite{Ali-Haimoud:2016mbv} obtain a semi-analytical solution for this system of equations, yielding the density, velocity and temperature profile vs. radius $r$ for a broad range of parameters $\beta_{\rm drag}$ and $\beta_{\rm cool}$. These reduce to the Bondi solution (where at small $r$ one finds $\rho\propto r^{-3/2}$ and $T\propto r^{-1}$) in the limit $\beta_{\rm drag},\, \beta_{\rm cool}\to 0$. 

Concerning radiative emission near the Schwarzschild radius, in the unmagnetised (or poorly magnetised) cosmological environment, it is  dominated by Bremsstrahlung radiation, i.e. free-free emission in a fully-ionised thermal electron-proton plasma, where densities obey $n_e=n_p$ and the common temperature is $T$. The expression for the energy-integrated emissivity $j$ is consistent with a simple estimate, writing 
 \begin{equation}
j=n_e^2 \alpha  \sigma_T T \mathcal{J}(X)
 \end{equation}
 where $X\equiv T/m_e$ and $J(X)$ is a dimensionless function (typically of order $\sim 10$) which can be expressed within a few percent accuracy by the fit~\cite{Ali-Haimoud:2016mbv}
\barr
\mathcal{J}(X) \approx 
\begin{cases}
\frac4{\pi} \sqrt{2/\pi} X^{-1/2} \left(1 + 5.5 X^{1.25} \right),  & X < 1,\\[6pt]
\frac{27}{2 \pi} \left[\ln(2 X {\rm e}^{- \gamma_{\rm E}} + 0.08) + \frac43\right],  & X > 1,~~~~~
\end{cases}
\earr
 including both $e-e$ and $e-p$ processes. Also, despite the problem being in general a complicated radiative transfer one, the authors of~\cite{Ali-Haimoud:2016mbv} bracket the efficiency parameter $\epsilon$ in the two limiting cases where ionisation of the matter either proceeds via collisional effects (if the emerging radiation field is too weak to photoionise the gas) or, in the opposite limit, via photoionisation. In both cases, one has
\beq
\epsilon \approx  \alpha \frac{ \dot{M}}{L_{\rm Edd}}\frac{T_{\rm S} }{m_p }\mathcal{J}\left(\frac{T_{\rm S}}{m_e}\right).\label{epscosmosph}
\eeq
where $\alpha$ is the fine-structure constant and $T_S$ the temperature at the Schwarzschild radius, predicted to be between 10$^9$ and 10$^{11}$ K for the collisional and photo-ionisation case, respectively. Apart for factors of order unity, due for instance to neglecting the effect of the Helium fraction or general-relativistic corrections,  the solution to this simplest formalisation of the cosmological problem is bracketed between two extremes.

\subsection{Consistency checks and domains of validity}\label{consistency}
The previous solution is a steady-state one, but the universe evolves over the dynamical timescale $H^{-1}$. As long as the  characteristic timescale of the accretion system is shorter than the cosmological timescale, the steady-state solution is meaningful. For that, one must require~\cite{Ricotti:2007jk}
 \begin{equation}
\frac{r_B}{c_{s,\infty}}H(z)<1 \Longrightarrow M\lesssim 10^{4.5}\,M_\odot\,,
 \end{equation}

Also, the calculation assumes that gas accretes on an isolated BH, which is valid as long as the Bondi radius is much smaller than the characteristic proper separation between PBHs, yielding~\cite{Ali-Haimoud:2016mbv}
 \begin{equation}
 M\lesssim 3\times 10^{4}f_{\rm PBH}^{-1/2}\,M_\odot\,,
 \end{equation}
This condition is only violated if the PBH are subject to strong clustering. Qualitatively, neglecting clustering (as henceforth assumed) leads however to more conservative constraints, since accretion grows quadratically with the mass.

In obtaining cosmological bounds, one implicitly considers homogeneous effects of the PBH on the medium, i.e. $x_e(z)$ is not space-dependent. This makes sense as long as each PBH can ionise all of the region separating it from the nearest PBH, leading to (Appendix A in~\cite{Serpico:2020ehh}) 
 \begin{equation}
f_{\rm PBH}> 10^{-15}x_e^3\frac{M_\odot}{M}\,.
 \end{equation}

Notice that we can use mean results ignoring the discreteness effect due to the finite number of PBHs in each patch of the CMB sky being analysed, as long as there is significantly more than a PBH in each patch of the CMB. The most stringent constraint comes from the maximum multipole used ($\ell\sim$2000), yielding~\cite{Serpico:2020ehh}
 \begin{equation}
N_{\rm PBH}\simeq \frac{5\times 10^7}{ \ell} \left(\frac{f_{\rm PBH}M_\odot}{M}\right)^{1/3}> 1\,.
 \end{equation}

Also note that it was explicitly checked in~\cite{Ricotti:2007au,Ali-Haimoud:2016mbv} that the plasma is optically thin to both Compton scattering and free-free absorption, as long as the accretion-rate is sub-Eddington. This hypothesis is used in obtaining the estimate of eq.~\eqref{epscosmosph}. The validity of the spherical accretion regime for such moderate accretion rates was also defended in~\cite{Ricotti:2007au}.

\subsection{Challenges}
If the above considerations provided a realistic and complete description of the accretion phenomenon in the cosmological framework, the robustness of the derived constraints would not be so  debated in the literature. In the following, we discuss some of the subtle points omitted in the previous treatment, and still raising questions.

First of all, the solution derived assumes PBHs at rest in the cosmological frame. Ref~\cite{Ali-Haimoud:2016mbv} generalised it to 
a finite PBH velocity $v_{\rm PBH}$, by replacing $c_{s,\infty}$ with  $v_{\rm eff}=\sqrt{c_{s,\infty}^2+v_{\rm PBH}^2}$ and $T_\infty\to T_\infty +m_pv_{\rm PBH}^2/(1+x_e)$. This prescription was also followed in~\cite{Ricotti:2007au}, and it is  inspired by what discussed already by Bondi~\cite{Bondi:1952ni}, as mentioned after eq.~\eqref{genAcc}. 
There is no strong argument supporting the consistency or the reasonable nature of  this recipe, as already noted in~\cite{Ali-Haimoud:2016mbv}. Of course, this would be only of academic  relevance if relative velocities between PBHs and the gas satisfied $v_{\rm PBH}\ll c_{s,\infty}$, but this is not what assumed in~\cite{Ali-Haimoud:2016mbv}. The rationale for supersonic velocities has to do with an effect discussed in~\cite{Tseliakhovich:2010bj}: At the recombination time, the sound speed in the baryonic fluid drops from relativistic to the thermal velocities of the hydrogen atoms,
becoming less than the relative velocity of baryons with respect to DM computed via linear perturbation theory.  Supersonic coherent flows of the baryons  are expected to exist (at cosmological scales) with respect to the underlying DM potential wells and thus PBHs, as long as PBHs trace the cold DM fluid. 

However, it has been questioned that such a simple picture catches the whole story. In~\cite{Kashlinsky:2020ial}, it has been argued that applying the reasoning within $\Lambda$CDM of ref.~\cite{Tseliakhovich:2010bj} to a PBH cosmology is incorrect at small scales, where the PBH Poisson clustering  is responsible for quickly driving the relative velocity of PBH and baryons to zero at the small scales relevant for accretion. 

Another caveat was raised in~\cite{Poulin:2017bwe} and concerns the validity of the spherical approximation, at least for the innermost accretion regions responsible for the bulk of the luminosity. For accretion to proceed spherically all the way down to the ISCO, the angular momentum of the fluid must be very small, smaller than expected by typical fluctuations in density at the accretion distance due to the small-scale power spectrum due to PBH, which is enhanced compared to $\Lambda$CDM. 

Both effects just mentioned are more and more relevant, the higher $f_{\rm PBH}$ is. In the opposite limit of small $f_{\rm PBH}$, one may be reduced to a situation where the solution of~\cite{Tseliakhovich:2010bj} applies: Yet, one should then consider that when the relative motion between a BH and the gas is supersonic,  an accretion disk is typically formed, as confirmed by simulations (see e.g.~\cite{Park:2012cr}). 

Overall, we expect that  when characteristic velocities between baryons and PBHs are of the order of $c_{s,\infty}$ and the universe is as homogeneous as expected in $\Lambda$CDM at small scales,  the spherical solution approximation of~\cite{Ali-Haimoud:2016mbv} should be applicable, modulo the replacement of $v_{\rm eff}$ with $c_{s,\infty}$. This  would result in bounds roughly one order of magnitude stronger than what discussed in~\cite{Ali-Haimoud:2016mbv}. In the opposite case, whenever baryonic angular momentum (e.g. due to gradients in density and velocities) and/or supersonic motions are relevant, we expect a disk to form in the inner region around the PBH, and the luminosity estimate would change.   We quickly review the physics relevant  to this case in Sec.~\ref{Disks}.

The study in~\cite{Ali-Haimoud:2016mbv} is also conservative in another respect. When $f_{\rm PBH}\ll 1$, another component is present in the universe besides PBHs and baryons, some cold  fluid responsible for the bulk of the DM. The PBH further attracts this DM, which in turn boosts the PBH capability to attract baryons. A first treatment of this effect was attempted in~\cite{Ricotti:2007au}; a more modern study, both analytical and numerical, was presented in~\cite{Serpico:2020ehh}. The inclusion of this effect is dealt with in Sec.~\ref{DMhalo}.

\section{Accretion disks}\label{Disks}
\begin{figure}
\centering
\includegraphics[scale=0.5]{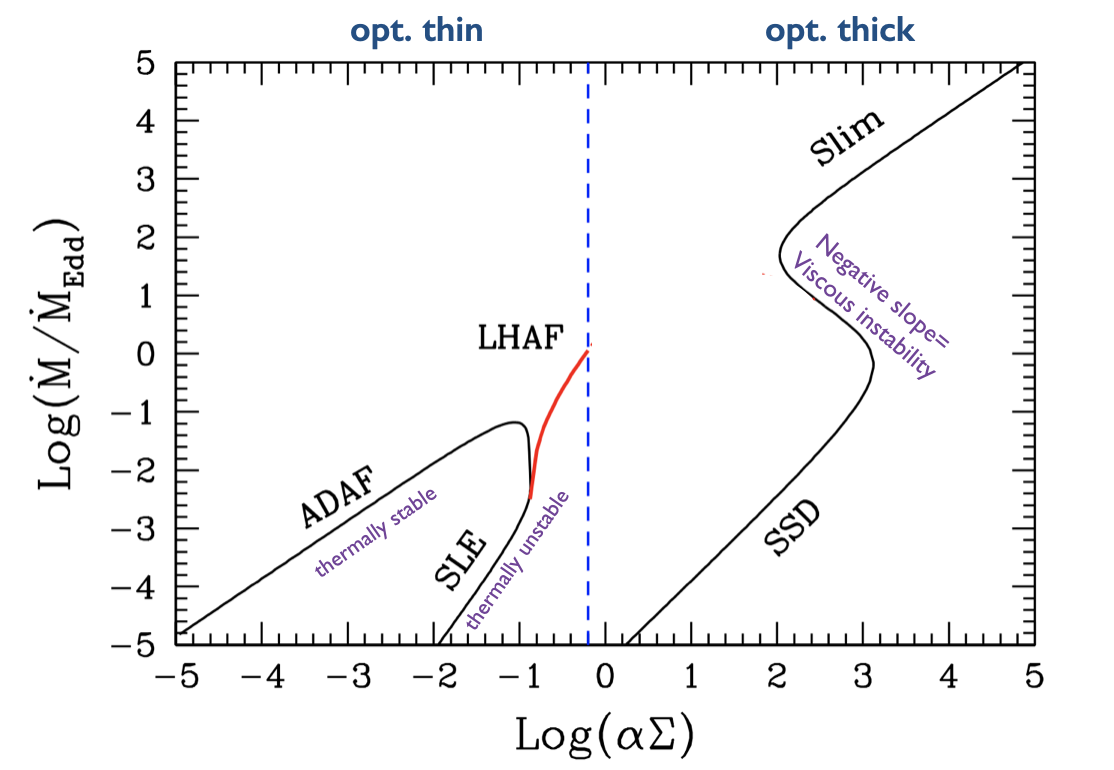}
\caption{A broad classifications of different regimes matching known accretion disk solutions, adapted from Ref.~\cite{Yuan:2014gma} by the author.} \label{fig:dzoo}
\end{figure}

Unfortunately,  no complete theory of disk accretion exists from first principles. 
A rather general classifications of different regimes matching several states seen in Nature has however been achieved, as summarised in Fig.~\ref{fig:dzoo}, adapted from Ref.~\cite{Yuan:2014gma}.
The two key parameters controlling most of the physics are the accretion rate ($y$ axis in  Fig.~\ref{fig:dzoo}, normalised to $\dot M_{\rm Edd}\equiv 10 L_{\rm Edd}$) and the optical depth (for which a proxy, the vertically integrated surface density $\Sigma$ in units of g/cm$^2$ times the dimensionless viscosity parameter $\alpha$, is reported in the $x$-axis of Fig.~\ref{fig:dzoo}).
 
If the disk is {\it optically thick}, the associated solutions are known as {\it cold flows}, since associated with relatively low temperatures: Baryons and electrons thermalise,  attaining a  radial-dependent temperature $T_c$ which  can be deduced by equating the dissipation rate per unit face area of the disk to the blackbody flux. One infers the figure of merit (e.g. see eq. 1.10 in~\cite{2002apa..book.....F}) 
\beq
T_c\simeq 4\times 10^7 {\rm K}\left(\frac{L}{L_{\rm Edd}}\right)^{1/4}\left(\frac{M}{M_\odot}\right)^{1/4}\sqrt{\frac{3\,{\rm km}}{R}}\,.\label{Tcold}
\eeq
Typically, the hottest regions of stellar mass BH disks in this regime do not exceed $\sim10^7$ K. If the accretion rate is low or moderate, in the optically thick regime the standard Shakura-Sunyaev solution~\cite{Shakura:1972te} is applicable (labelled SSD in Fig.~\ref{fig:dzoo}). Its key assumption is that the dissipation happens through viscous effects parametrised by the dimensionless coefficient $\alpha\lesssim 1$, independent of other properties of the disk, so that viscous torque is proportional to the pressure. In this solution, the disk mass is negligible compared to that of the accreting BH and  self-gravity of the disk can be neglected. The efficiency parameter  $\epsilon\simeq 0.1$ is associated to this solution and adopted in most related literature. 

This solution breaks down if accretion grows to values comparable to the Eddington scale. Other stable solutions exist at high accretion rate and high optical thickness. When the accreting gas becomes too optically  thick to radiate all the  locally dissipated energy, radiation is trapped and advected inward with the accretion flow.  The radiative efficiency becomes lower than ~10\% and, as a consequence, $L$ never significantly exceed $L_{\rm Edd}$. This alternative regime of the cold flow class is sometimes called {\it slim disk} (since one can still deal with is via vertically integrated equations, as in the case of the solution~\cite{Shakura:1972te}) but it is better characterised as {\it optically thick advection-dominated accretion flow}. 

In the opposite regime of {\it optically thin disk}, thermal equilibrium is not achieved and the corresponding solutions are classified as {\it hot flows}, since the relevant temperature scale is the virial one,
\beq
T_{\rm vir}\simeq \frac{GM m_p}{3R}\simeq \frac{m_p}{6} \frac{R_S}{R}\simeq 1.7\times 10^{12} {\rm K}\frac{M}{M_\odot}\frac{R_S}{R}\,,
\eeq
clearly satisfying the relation $T_{\rm vir}\gg T_c$. The state of the plasma is typically parameterised via two temperatures, one for leptons $T_e$ and one for nuclei $T_{\rm ion}$, satisfying $T_{\rm ion}\simeq T_{\rm vir}\gg T_{e}$ (The first such model, dubbed SLE in Fig.~\ref{fig:dzoo}, was proposed in 1976 in~\cite{1976ApJ...204..187S}, but was realised to be thermally unstable.) In general, under the described conditions the accreting plasma forms a thick torus, and a sizeable fraction of accretion energy goes into heating the flow rather than being radiated away.  This regime is known as {\it optically thin advection-dominated accretion flow}, or  ADAF in short (bottom-left in Fig.~\ref{fig:dzoo}).  This regime is characterised by an efficiency $\epsilon\simeq 0.1f(\dot M)$, with $f\lesssim 1$ and $f'>0$; $\epsilon$ saturates to 0.1 at large accretion rates, i.e. when approaching Eddington condition, a regime sometimes dubbed luminous hot accretion flows (LHAF in Fig.~\ref{fig:dzoo}).
 The radiative efficiency of ADAF largely depends on  the fraction of energy shared by the electrons, $\delta$.  According to~\cite{Yuan:2014gma}, theoretical lower limits indicate $\delta\gg 0.01$, and fits to  observations suggest $0.1<\delta<0.5$. For conditions where multiple stable accretion regimes can coexist, observations seem to support the conclusion that Nature selects a ADAF-type solution (``Strong ADAF principle'').

Some effects have been neglected in the above schematic classification. Notably,  dynamical effects due to magnetic fields, the BH spin, radiative feedback and kinetic (as opposed to fluid) effects have been neglected. While this is still an open field of research, the most important phenomenological correction to the above picture relates to the presence of outflows and jets: Some of the material falling on the BH can be actually carried away from the inner disk in the form of a non-relativistic wind or a relativistic jet, either Poynting-flux or matter dominated. In extreme cases, so-called MAD (magnetically arrested disk) solutions apply. For observed active galactic nuclei (most often in the ADAF regime) the role of these outflows is relevant, suppressing the BH-associated luminosity by one order of magnitude or more compared to what expected from the Bondi solution~\cite{Pellegrini:2005pi}. These effects seem to be particularly important in quiet BH, such as Sgr A* at the center of the Milky Way.  In the light of this situation, a rather conservative approach was taken in works like~\cite{Poulin:2017bwe,Serpico:2020ehh} when dealing with a disk model in a cosmological context: Only the low-efficiency ADAF models were considered, with a choice of $\delta=0.1$ for the energy shared by electrons (roughly the minimum suggested value consistent with observations according to~\cite{Yuan:2014gma})  and a suppressed accretion with respect to the (cosmological) Bondi-like solution, to mimic the effects of the outflows in a way roughly consistent with observations~\cite{Pellegrini:2005pi}. 

That said, the need to go beyond the simple ADAF models in the cosmological setting and for deriving CMB anisotropy bounds is questionable, and it may well be that the prescriptions adopted in~\cite{Poulin:2017bwe,Serpico:2020ehh} are overly conservative. First of all,  PBHs in a cosmological setting are very different from astrophysical ones at low $z$: Models typically predict them to form preferentially with vanishing spin and they live in an unmagnetised medium. To mention but one important element, jets are typically Poynting flux dominated are are associated to magnetic effects as well as the BH spin. We expect these effects to be very small if not absent in the cosmological context.   Also, we cannot emphasise enough that  {\it energy is not lost}: even outflow and jet luminosities, if present, should eventually cause heating and excitation/ionisation of the cosmological medium, hence contributing  to the CMB bounds even when not contributing to the BH luminosity proper. This remark is also relevant to the recent trend of applying simulation results such as~\cite{Park:2012cr}, showing important {\it radiative feedback}, to the problem of cosmological bounds on PBHs, see e.g~\cite{Facchinetti:2022kbg,Agius:2024ecw}. This feedback effect consists in the modification to the actual accretion rate onto the BH due to the (sizeable) modification of the medium surrounding the BH brought upon by the accretion luminosity itself, which in turn also induces a time-dependence.  However, even leaving apart the reliability of the extrapolation of the medium density/temperature parameters of~\cite{Park:2012cr} to cosmological values, Refs.~\cite{Facchinetti:2022kbg,Agius:2024ecw} only consider the effects of the decreased UV/X-ray luminosity associated to the BH. Whenever significant feedback is suspected, by definition a significant fraction of the kinetic energy acquired by the baryons ``falling from infinity'' does not disappear into the BH, but stays in the surrounding environment : either directly (as heating of the gas that never penetrates close to the BH) or indirectly, e.g. by contributing to the ionisation of a large Str\"omgren bubble surrounding the BH or the light it releases in the medium. While it is understandable that the energy budget associated to those is ignored in the simulations of~\cite{Park:2012cr}, which are primarily meant to be applied to the UV/X-ray emission of astrophysical BH, they should be relevant (and at least discussed) for CMB constraint applications. Currently, however, they are simply ignored~\cite{Facchinetti:2022kbg,Agius:2024ecw}.

Till now, we have glossed over the spectral features of the emission. In general, modern tools like \texttt{EXOCLASS} allow one to reliably compute the energy deposited in the cosmological medium, once the injected spectrum is known. As confirmed also by analytical approximations (see notably~\cite{Ali-Haimoud:2016mbv}), at least in the most relevant redshift range $300\lesssim z\lesssim 1000$, most of the injected energy is absorbed by the medium; its repartition in ionisation, excitation, heating is only weakly dependent from the details of the spectrum. The latter can be computed rather precisely within the simplifying assumptions of~\cite{Ali-Haimoud:2016mbv}, so that the uncertainty on the bounds related to the spectrum is definitely not the dominant one. 

Even for disk models, the situation is qualitatively similar. As reported in~\cite{Yuan:2014gma}, the dominant electron emission in the context of ADAF can be parameterised as (see~\cite{Poulin:2017bwe})
\beq
\frac{{\rm d} L}{{\rm d}{E}}\propto\Theta(E-E_{\rm min})E^{-a}\exp(-E/T_s)\,,
\eeq
with $a\simeq 0-0.5$, $T_s\simeq m_e$ and $E_{\rm min}\simeq 10\,$eV, accounting for a cut including only the fraction of the spectrum ``useful'' for excitation/ionisation. While details do depend on the accretion rate and BH mass,  due to the above-mentioned weak dependence of the deposited energy on the spectral shape, these variations only lead to a moderate spread and uncertainty in the derived bounds, quantified  in a factor $\simeq 2$ in ref.~\cite{Poulin:2017bwe}.

\section{Including dark matter accretion}\label{DMhalo}

If PBHs {\it do not} constitute the totality of the DM, besides baryons they also gravitationally attract DM~\cite{Bertschinger:1985pd,Mack:2006gz,Berezinsky:2013fxa}. In absence of energy and angular momentum loss, as expected for most DM candidates, most of this DM does not accrete the PBH, rather forms a halo, which can however boost the PBH capability to accrete baryons~\footnote{One implication is that the $\dot M$ associated to both baryons and DM is insufficient, once integrated down to  $z\sim 100$, to significantly alter $M$. For our purposes, we can consider $f_{\rm PBH}\simeq const.$}. 
Although the original Bondi problem was considering accretion onto a point particle, a natural generalization of the notion of Bondi radius for an extended distribution of mass can be written as~\cite{Park:2015yrb}:
\begin{equation} \frac{G_N\,M_{\rm PBH}}{r_{\rm B,eff}}-\Phi_h(M_{\rm PBH},r_{B,{\rm eff}},t)=v_{\rm eff}^2(t)\,,\label{effRB}
\end{equation}
where $M_{\rm PBH}$ is the initial PBH mass and $\Phi_h$ the (time-dependent) gravitational potential associated to the DM halo.
In~\cite{Serpico:2020ehh},  the problem was thus tackled by adopting Eq.~(\ref{HLAcc}), with $r_{\rm B}$ replaced by $r_{\rm B,eff}$ solution of Eq.~(\ref{effRB}), with the gravitational potential of the halo estimated analytically or numerically.  

Analytically, one can consider the spherically symmetric case of growth of a halo of (exactly cold and dispersionless) DM, around a  PBH which is the only center of attraction in the whole universe. The time evolution of  a mass-shell at position $r$ can be obtained by solving the following differential equation,
\begin{eqnarray}
  \label{eq:d2rdt21}
  \frac{{\rm d}^2r}{{\rm d}t^2} =-\frac{4 G_{\rm N}} \pi{3}
  r \left[\rho_{\rm PBH}+  \sum_i 
\left( \rho_i + 3 p_i \right)\right],
\end{eqnarray}
where $\rho_i$ and $p_i$ are the energy density and pressure of the component ``$i$'' (radiation, matter, and dark energy, if effective at the redshift considered), respectively, and we defined the energy density of the PBH as $\rho_{\rm PBH} = 3\,M_{\rm PBH}/(4 \pi r^3)$.   The
physical radius $r$ is represented by $r= a(t)x$ where $a(t)$ is the scale factor normalised as $a_0 = 1$ at present, and $x$ is the co-moving coordinate. 
At each time $t$, the DM halo mass is the DM density integrated up to the radius $r_s$ defined by 
${\rm d} r_s/{\rm d} t=0\,$. 
This approach yields the following key results~\cite{Serpico:2020ehh,Mack:2006gz}:
\begin{itemize}
    \item A time evolution given by
\begin{eqnarray}
  \label{eq:scalingLawCorr}
  M_{\rm halo} \simeq \left( \frac{3000}{1 + z} \right)  M_{\rm PBH}\,.
\end{eqnarray}
\item A density profile proportional to $\propto r^{-3}$ down to small distances, where a free-fall profile $r^{-3/2}$ takes over.
\end{itemize} 

Eq.~(\ref{eq:scalingLawCorr}) only yields an upper limit to the  DM halo mass. At least at late times, the growth breaks down, e.g. due to tidal effects of nearby PBHs and other halos. It is also worth noting that the radial profile crucially depends on 
the free-fall boundary condition at the center; not accounting for the DM  angular momentum is expected to be a too crude approximation.
Resorting to numerical simulations,  in~\cite{Serpico:2020ehh} it was found that as long as $f_{\rm PBH}$ is not too large and the redshift is not too low (evolution was followed down to $z=99$), halos accrete onto isolated PBHs, and the profile is independent of $f_{\rm PBH}$.  Over about two decades in radius, at early times the profile matches the $r^{-2.25}$ power-law predicted by~\cite{Bertschinger:1985pd}, confirming the conjecture advanced in Ref.~\cite{Ricotti:2007jk}, Sec. 4; at late times, the profile evolves towards the slightly steeper $r^{-2.5}$.  This power-law profile is consistent with those found in independent numerical simulations by \cite{Adamek:2019gns}, which also found a smooth transition to standard NFW-like profile of DM halos at large radii.  The Poisson's equation for the gravitational potential $\phi_{i} = (4\pi G)^{-1} \nabla^{-2} \rho_i$
is solved in Fourier space  for PBHs and DM separately.  The unknown $r_{\rm B,eff}$ follows from plugging the potential thus found in Eq.~(\ref{effRB}). 

At high redshifts, the PBH is much more relevant than the DM halo and $r_{\rm B,eff}$ is small, close to the naive estimate  $r_{\rm B}$ for the naked PBH. When the halo eventually exceed the PBH mass, its further evolution is largely independent of the PBH contribution to the potential. The numerical results lead to an estimated halo mass which is about 60\% of the simple result of eq.~\eqref{eq:scalingLawCorr}, with a similar scaling with redshift, although with a different mass profile. Note that it is safe to neglect the ``ordinary'' DM halos feedback onto the halos growing around PBH, since the former ones only grow at much later times (typically $z\lesssim 30$ in a $\Lambda$CDM cosmology) than those of concern for us. {\it A fortiori}, the feedback of the baryons can also be neglected.  Also, note that most baryons are still unbound to halos, and their ratio to the DM in the growing halos around PBH is much smaller than the baryon to DM cosmological density ratio of $\sim 15\%$. Hence, objections on the realism of power law DM density profiles  around BH surviving in the current universe~\cite{Ullio:2001fb} do not apply to the pristine configurations considered here.
 
A simple semi-analytical model  that matches these results and interpolates between limiting behaviours was proposed in~\cite{Serpico:2020ehh}.
In the specific case of a point-like potential due to the PBH plus  the power-law matter distribution around it, with density $\rho(r)\propto r^{-\nu}$ up to a distance $r_h$ and total mass $M_h$, Eq.~(\ref{effRB}) rewrites
\begin{equation}
\frac{ v_{\rm eff}^2(z)r_{B,{\rm eff}}}{G_N\,M_{\rm PBH}}=1+ \frac{M_h}{M_{\rm PBH}}\left\{\Theta(r_{B,{\rm eff}}-r_h)+
\frac{\Theta(r_h-r_{B,{\rm eff}})}{1-p}\left[\left(\frac{r_b}{r_h}\right)^p-p\left(\frac{r_{B,{\rm eff}}}{r_h}\right)\right]\right\}\,,\label{effbondi}
\end{equation}
where $p\equiv 3-\nu$, and $M_h$ and $r_h$ depend from $M_{\rm PBH}$ and $z$. Ref.~\cite{Serpico:2020ehh} adopted eq.~\eqref{eq:scalingLawCorr} for  the halo mass $M_h$, assuming it is extending up to the turnaround radius,
defined as (see e.g. Sec. 4 in Ref.~\cite{Ricotti:2007jk})
\begin{equation} 
r_{h}\simeq 58\, {\rm pc}\,(1+z)^{-1}\left(\frac{M_h(M_{\rm PBH},z)}{M_\odot}\right)^{1/3}\,.
\end{equation}

Eq.~\eqref{effbondi} admits either the solution (``Bondi radius'' associated to the halo mass)
\begin{equation}\label{smallhalo}
r_{\rm B,eff}=\frac{G_N(M_{\rm PBH}+M_{h})}{v_{\rm eff}^2}\simeq  \frac{G_N\,M_{h}}{v_{\rm eff}^2}\equiv r
_{{\rm B},h}\,,
\end{equation}
which holds if $r_h<r_{{\rm B},h}$; otherwise, if $r_h> r_{{\rm B},h}$, neglecting the PBH mass one has
\begin{equation}\label{largehalo}
r_{\rm B,eff}\simeq r_h
\left[(1-p)\frac{r_h}{r_{{\rm B},h}}+p\right]^{\frac{1}{p-1}}\,\leq r_h.
\end{equation}
Note that Eq.~(\ref{largehalo}) tends to $r_{{\rm B},h}$ when $p\to 0$, as expected: When the DM halo profile is very steep and/or the halo is very compact, as far as accreting baryons are concerned they simply see a BH whose effective mass is the sum of the PBH and the DM halo mass. If the halo is fluffy or large, only a fraction of the mass of the halo contributes to the accretion.
In any case, the condition $r_{{\rm B, eff}}\geq r_{\rm B, PBH}$ must hold. In Ref.~\cite{Serpico:2020ehh}, this model with the choice $p=0.75$ suggested by simulations was thus used to asses  the impact of PBH accretion onto the CMB.

\section{Constraints and perspectives}\label{ConstrDisc}
\begin{figure}
\includegraphics[scale=0.34]{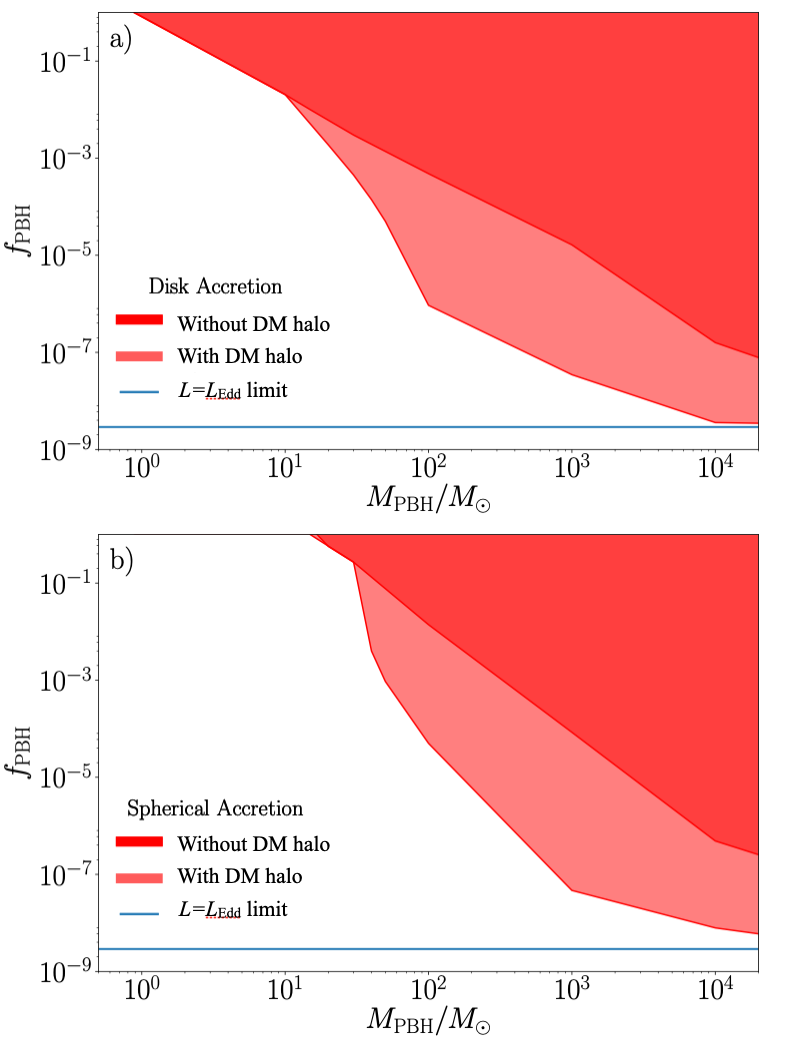}
\caption{Bounds on the abundance of PBH assuming disk accretion (top panel) or spherical accretion (bottom panel). We show the results with (light-shaded) and without (dark-shaded) the formation of a DM halo. The horizontal line shows the limiting bound assuming constant emission at Eddington luminosity. Adapted from Ref.~\cite{Serpico:2020ehh}; see text for details.}
\label{fig:bounds}
\end{figure}

In the limit of a monochromatic mass distribution for PBHs, fig.~\ref{fig:bounds} (from~\cite{Serpico:2020ehh}) reports the bound for the ``conservative' ADAF disk model described in Sec.~\ref{Disks} (top panel), and the spherical accretion model of~\cite{Ali-Haimoud:2016mbv} (with $v_{\rm eff}=c_{s,\infty}$) described in Sec.~\ref{AccCosmo} (bottom panel).  In each case, light shaded regions represent the excluded parameter space when including the DM halo effect. 
Here are the datasets from which these bounds are derived: {\em Planck} 2018 high-$\ell$ and low-$\ell$ TT, EE and lensing likelihood \cite{Planck:2018vyg,Planck:2019nip}; the isotropic BAO measurements from 6dFGS at $z = 0.106$~\cite{Beutler:2011hx}  and from the MGS galaxy sample of SDSS at $z = 0.15$~\cite{Ross:2014qpa}; the anisotropic BAO and the growth function $f\sigma_8(z)$ measurements from the CMASS and LOWZ galaxy samples of BOSS DR12 at $z = 0.38$, $0.51$, and $0.61$~\cite{Alam:2016hwk};  the Pantheon supernovae dataset \cite{Scolnic:2017caz}, which includes measurements of the luminosity distances of 1048 SNe Ia in the redshift range $0.01 < z < 2.3$.  The non-CMB datasets are essentially instrumental in breaking some degeneracies. For technical details on the derivation of the bounds and the analysis chain, see~\cite{Serpico:2020ehh}. 
A reasonably conservative statement is that the totality of DM in the form of PBH (i.e. $f_{\rm PBH}=1$) is excluded for $M\gtrsim 4\,M_\odot$, with a factor $\sim 5$ systematic uncertainty dominated by the poorly known accretion luminosity.  The constraints on $f_{\rm PBH}$ improve more than linearly with mass up to hundreds of solar masses: The scaling derived in~\cite{Poulin:2017bwe} with the mass as well as the accretion eigenvalue was $f_{\rm PBH}^{\rm lim}\propto (M\,\lambda_{\rm eff})^{-1.6}$. The dependence is stronger if the DM halo effect is taken into account, as it should. At least from $\sim 50\,M_\odot$ to $\sim 10^3\,M_\odot$, the CMB anisotropy bounds are nominally the tightest known to date. If extrapolated to high masses, they saturate eventually at $f_{\rm PBH}\simeq 3\times 10^{-9}$ around $10^4\,M_\odot$, when the luminosity is expected to approach the Eddington one. The reliability of the bounds in this range is however doubtful, since the accretion physics extrapolation (see Sec.s~\ref{AccCosmo}, \ref{Disks}) as well as some of the cosmological assumptions (see Sec.~\ref{consistency}) break down. 

Further, there is no indication that assuming a monochromatic mass distribution leads to over-conservative bounds. Matching the trend typically observed in other cases where a broader mass function is used~\cite{Kuhnel:2017pwq,Carr:2017jsz}, also CMB bounds are stable or improve for broad mass functions. In~\cite{Poulin:2017bwe}, by recasting the monochromatic bounds according to the technique suggested in~\cite{Carr:2017jsz},
it was checked that the constraints derived on the monochromatic mass function also apply to the average mass value of a broad, log-normal mass distribution; the broad distribution is more tightly constrained when its width covers a decade or more in mass space. More recently, in~\cite{Juan:2022mir}, the CMB anisotropy bound was also computed for a very broad and multi-peaked mass function $\psi(M)$. This was done by ``brute force'',  integrating numerically the bounds obtained for each mass value, i.e. by using deposition energies  
\begin{equation}
\frac{{\rm d}E}{{\rm d}V{\rm d}t}(z)\Bigg|_{\rm dep, c}  = \int \dd M f_c (z, M) \psi (M) \frac{{\rm d}E}{{\rm d}V{\rm d}t}\Bigg|_{\rm inj} ,
 \label{energy_dep}
\end{equation}
 or by recasting the monochromatic bounds, as done in~\cite{Poulin:2017bwe}. For the case considered, the ``exact'' numerical bounds for a very broad mass function are about 50\% more stringent than if approximately computed via the prescription of~\cite{Carr:2017jsz}. 

The most important phenomenological consequence of these bounds is to provide an independent argument against the possibility that the bulk of the surprisingly heavy BH population (a few tens $M_\odot$ to $\sim 100 M_\odot$) inferred by LIGO and Virgo constitutes a significant fraction of the DM. However, the possibility that some of these objects is of primordial origin is still marginally consistent with the CMB anisotropy bounds, at least for some assumptions on accretion. As a side remark, despite the tight bounds at large masses and the fact that some assumptions become shakier in this range (See Sec.~\ref{consistency}), the CMB angular power spectra do not exclude a primordial origin hypothesis for the SMBH, if combined with a significant accretion phase, although this hypothesis is challenged by other arguments (see Sec. V.B in~\cite{Serpico:2020ehh} as well as ref.\cite{Juan:2022mir}) and no convincing self-consistent model has been proposed,  yet.

The major uncertainties still plaguing these constraints are not the statistical errors on the data, rather the theory systematics related to accretion physics, as discussed at length. Based on current knowledge, the bounds discussed here should rather err on the conservative side. Also, we have only considered PBH accretion in the dark ages, stopping integration at $z\sim 100$, where cosmological linear perturbation theory is reliable. Accretion certainly continues once halos become self-gravitating and capable of retaining significant amount of baryons, but extrapolation of the cosmological solutions discussed here cannot be considered educated guesses, since density, velocities, and accretion properties will be dominated by the astrophysical conditions in the (proto)halos.
It is worth keeping in mind, however, that if a significant growth of the PBH mass in the dark ages is excluded, this is not necessarily the case at lower redshift (although such a situation may raise more phenomenological problems than it solves). The bounds from CMB anisotropies must be implicitly thought as bounds on the primordial value of $f_{\rm PBH}$. 

The most important improvement in this field is expected from dedicated numerical simulations. Note that some recent analysis~\cite{Facchinetti:2022kbg} has attempted to directly use the results of hydrodynamical simulations, notably those of Ref.~\cite{Park:2012cr}, to describe a supersonically moving accreting BH. Reassuringly, for masses between 1 and 100 $M_\odot$, the fiducial model of Ref.~\cite{Park:2012cr} does lead to constraints bracketed by the spherical ones of~\cite{Ali-Haimoud:2016mbv} and the disk one of Ref.~\cite{Poulin:2017bwe}, obtaining even stronger constraints at $M\gtrsim 10^2\,M_\odot$~\footnote{The authors of Ref.~\cite{Facchinetti:2022kbg} also present results combining the accretion extracted from Ref.~\cite{Park:2012cr} 
with values for the radiative efficiency  taken from the analysis of Ref.~\cite{Ali-Haimoud:2016mbv}, yielding more conservative bounds. We note however that this choice does not match any result or prescription in the literature, and combines ingredients which appear mutually incompatible (for instance, the solution of Ref.~\cite{Ali-Haimoud:2016mbv} assumes spherical symmetry, while the configuration in Ref.~\cite{Park:2012cr}  leads to disk formation).}.  As discussed in Sec.~\ref{Disks}, however, the main limitation of these exercises is in accounting for all other channels of energy into which the initial, ``kinetic'' accretion luminosity is leaking into, which is inherent in the claim of significant feedback. Qualitatively, this should obviously improve the CMB bounds, but a quantitative analysis is currently missing. 

Some interesting avenue for the future would be to refine the model of~\cite{Ali-Haimoud:2016mbv} for steady-state, spherical accretion in  an unmagnetised  medium of cosmological density at rest with respect to the BH, but including radiative transfer  and possibly GR effects. This would collapse the results currently bracketed between the collisional and photoionisation regime into a well-defined benchmark family of solutions. 
Eventually, it would be important to extend those simulation to the case of  relative motion between the PBH and the gas (notably in the supersonic regime). The energy carried by eventual outflows and jets, if present, should also be estimated. Since the cosmological medium does not share a number of complications with the astrophysical medium (think of the lack of large inhomogeneities and magnetisation, vanishingly low metallicities, absence of dust grains, etc), a determination of universal laws of the accretion rate and luminosity could be within reach. 
From the cosmological theory side, clarifying the expectation for the power spectrum and the relative motion between PBHs and baryons at very small scales not only in $\Lambda$CDM, but also in a cosmology dominated by PBHs, is another intriguing direction for improvement.

\begin{acknowledgement}
I would like to thank all the colleagues who have helped me deepening this fascinating topic, at the interplay of cosmology, astrophysics and particle physics. Special thanks to my co-authors on this subject, F. Calore, S. Clesse, G. Franco Abell\'an, J. Iguaz, D. Inman, K. Kohri, and V. Poulin.
\end{acknowledgement}

 \bibliographystyle{unsrt}
 \bibliography{serpico.bib}

\end{document}